\documentclass[fleqn,12pt,twoside]{article}
\usepackage{espcrc1}

\usepackage{graphicx}

\newcommand{\AmS}{{\protect\the\textfont2
  A\kern-.1667em\lower.5ex\hbox{M}\kern-.125emS}}

\title{Probe Initial Parton Density and Formation Time via Jet Quenching}

\author{Xin-Nian Wang\address[MCSD]{Nuclear Science Division MS 70R0319\\ 
       Lawrence Berkeley National Laboratory \\ 
        Berkeley, California 94720}}
       
\begin{document}

\maketitle


\begin{abstract}
Medium modification of jet fragmentation function due to multiple scattering
and induced gluon radiation leads directly to jet quenching or
suppression of leading particle distribution from jet fragmentation.
One can extract an effective total parton energy loss which can be
related to the total transverse momentum broadening. For an expanding
medium, both are shown to be sensitive to the initial parton density
and formation time. Therefore, one can extract the initial parton
density and formation time from simultaneous measurements of parton
energy loss and transverse momentum broadening. Implication of the
recent experimental data on effects of detailed balance in parton
energy loss is also discussed.
\end{abstract}

\section{Introduction}
In high-energy heavy-ion collisions, one of the pressing issues is the
direct determination of the initial parton density of the dense medium 
and its formation time. 
Jet quenching or attenuation of leading particles from jet fragmentation 
can provide an effective tool to measure directly the
parton density of the medium with which the jet interacts strongly during
its propagation. Recent theoretical studies \cite{gw1,bdms,zhak,glv,wied}
have shown that parton energy loss via induced radiation is directly
related to the gluon density of the medium. The attenuation will 
suppress the final leading hadron distribution giving rise to
modified parton fragmentation functions \cite{wh}. By measuring
the medium modification of the fragmentation function one can thus
extract the effective parton energy loss. Most importantly, one can
compare the effective parton energy loss extracted from heavy-ion
collisions to that of cold nuclei and extract the initial parton density
of the hot dense medium relative to a cold nucleus.

\section{Parton Energy Loss in Cold Nuclei}

In deeply inelastic scattering (DIS) off a nuclear target, a quark 
suffers multiple scattering with other nucleons inside the nucleus after 
it was knocked out of its parent nucleon. The induced gluon bremsstrahlung 
before hadronization leads to suppression of leading hadrons and gives
rise to a modified quark fragmentation function. Including the leading 
twist-4 contributions from double scattering processes, the nuclear
medium induced correction to the fragmentation function is found to
be similar to the radiative correction in the vacuum except that the
normal splitting functions are replaced by the modified ones. These
modified splitting functions are found \cite{gw01} to depend on 
the quark-gluon correlation function $T^A_{qg}(x,x_L)/q_A(x)$. Here
\begin{eqnarray}
T^A_{qg}(x,x_L)&=& \int \frac{dy^{-}}{2\pi}\, dy_1^-dy_2^-
e^{i(x+x_L)p^+y^-+ix_Tp^+(y^-_1-y^-_2)}(1-e^{-ix_Lp^+y_2^-})
(1-e^{-ix_Lp^+(y^--y_1^-)}) \nonumber \\
&\frac{1}{2}&\langle A | \bar{\psi}_q(0)\,
\gamma^+\, F_{\sigma}^{\ +}(y_{2}^{-})\, F^{+\sigma}(y_1^{-})\,\psi_q(y^{-})
| A\rangle \theta(-y_2^-)\theta(y^--y_1^-)
\label{eq:qgmatrix}
\end{eqnarray}
and $q_A(x)$ is the twist-two quark distribution of the nucleus.
Such a two-parton correlation function contains essentially four 
independent twist-4 parton matrix elements 
in a nucleus [$x_T=\langle k_T^2\rangle/2p^+q^-z(1-z)$].
The dipole-like form-factor $(1-e^{-ix_Lp^+y_2^-})(1-e^{-ix_Lp^+(y^--y_1^-)})$
arises from the interference between the final state radiation of 
the $\gamma^*q$ scattering and the gluon bremsstrahlung induced
by the secondary quark-gluon scattering. By generalizing the 
factorization assumption to these twist-four
parton matrices, we have
\begin{equation}
T^A_{qg}(x,x_L)/q_A(x)\approx \widetilde{C}(Q^2)
m_NR_A (1-e^{-x_L^2/x_A^2}),
\label{eq:tqg2}
\end{equation}
with a Gaussian nuclear distribution 
$\rho(r)\sim \exp(-r^2/2R_A^2)$, $R_A=1.12 A^{1/3}$ fm.
Here, $x_A=1/m_NR_A$, and $m_N$ is the nucleon mass.

Since the two interference terms in the dipole-like form-factor
involve transferring momentum $x_Lp^+$ between different nucleons 
inside a nucleus, they should be suppressed for large nuclear size 
or large momentum fraction $x_L$. Notice that $\tau_f=1/x_Lp^+$ is 
the gluon's formation time. Thus, $x_L/x_A=L_A/\tau_f$, with $L_A=R_Am_N/p^+$ 
being the nuclear size in the infinite momentum frame.
The effective parton correlation and the induced gluon emission 
vanishes when the formation time is much larger than the nuclear size,
$x_L/x_A\ll 1$, because of the Landau-Pomeranchuck-Migdal (LPM) 
interference effect.
Therefore, the LPM interference restricts the radiated gluon to have 
a minimum transverse momentum $\ell_T^2\sim Q^2/m_NR_A\sim Q^2/A^{1/3}$. 
The nuclear corrections to the fragmentation function due to double 
parton scattering will then be in the order of 
$\alpha_s A^{1/3}/\ell_T^2 \sim \alpha_s A^{2/3}/Q^2$, which depends
quadratically on the nuclear size. For large values of
$A$ and $Q^2$, these corrections are leading; yet the requirement
$\ell_T^2\ll Q^2$ for the logarithmic approximation in deriving the 
modified fragmentation function is still valid.

\begin{figure}[htb]
\begin{minipage}[t]{80mm}
\includegraphics[width=82mm,height=62mm]{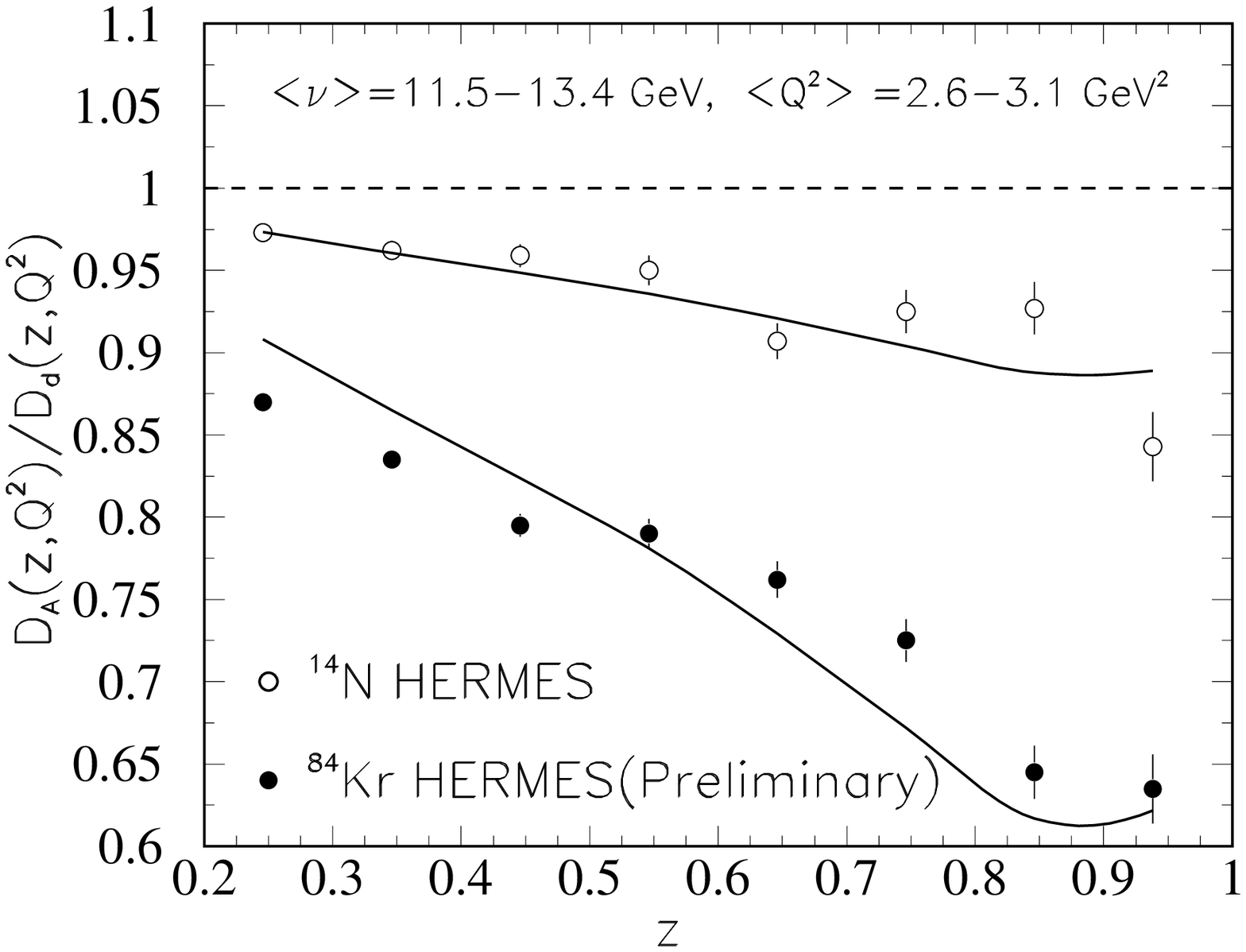}
\caption{Predicted nuclear modification of jet fragmentation function
is compared to the HERMES data \protect\cite{hermes} on ratios of
hadron distributions between $A$ and $D$ targets in DIS.}
\label{fig1}
\end{minipage}
\hspace{\fill}
\begin{minipage}[t]{75mm}
\includegraphics[width=82mm,height=62mm]{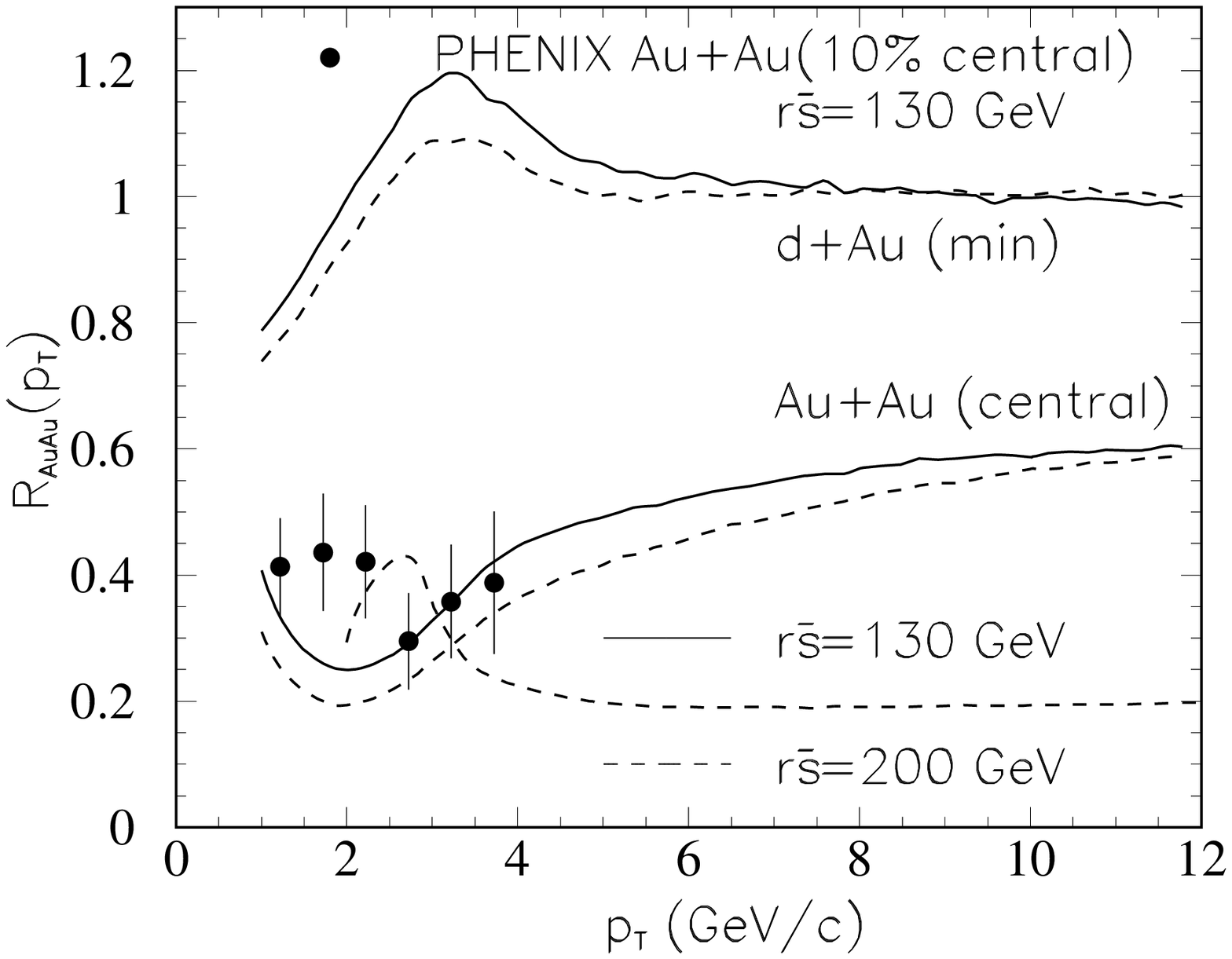}
\caption{Calculated nuclear modification factor 
compared to PHENIX data \protect\cite{phenix}.
The lower dashed line used the energy loss as given 
in Eq.~(\protect\ref{para}).}
\label{fig2}
\end{minipage}
\end{figure}

Shown in Fig.~\ref{fig1} are the calculated 
nuclear modification factor of the fragmentation functions for $^{14}N$ 
and $^{84}Kr$ targets as compared to the recent HERMES data \cite{hermes}.
The predicted shape of the $z$- and and energy dependence agrees well 
with the experimental data.  A remarkable feature of the prediction
is the quadratic $A^{2/3}$ nuclear size dependence, which is verified 
for the first time by an experiment. The only parameter in our
calculation is found to be $\widetilde{C}(Q^2)=0.0060$ GeV$^2$ 
with $\alpha_{\rm s}(Q^2)=0.33$ at $Q^2\approx 3$ GeV$^2$.

If one defines theoretically the quark energy loss as that carried
by the radiated gluons, then the averaged total energy loss is,
\begin{equation}
\Delta E=\nu\langle\Delta z_g\rangle
\approx  \widetilde{C}\alpha_{\rm s}^2(Q^2)
m_NR_A^2(C_A/N_c) 3\ln(1/2x_B).
 \label{eq:loss1}
\end{equation}
With the determined value of $\widetilde{C}$, 
$\langle x_B\rangle \approx 0.124$ in the HERMES experiment \cite{hermes}
and the average distance $\langle L_A\rangle=R_A\sqrt{2/\pi}$
for the assumed Gaussian nuclear distribution,one gets the quark energy 
loss $dE/dL\approx 0.5$ GeV/fm inside a $Au$ nucleus.

One can also calculate the transverse momentum broadening of the quark
jet which is also related to a similar twist-four parton matrix elements.
\begin{equation}
\langle \Delta q_\perp^2\rangle=\frac{2\pi\alpha_{\rm s}}{N_c}
\frac{\widetilde{T}^A_{qg}(x_B)}{q_A(x_B)}
\approx \widetilde{C}\frac{\pi\alpha_{\rm s}}{N_c}m_N R_A
\approx 0.011 A^{1/3} \rm{GeV^2}
\end{equation}

\section{Initial Parton Density in $Au+Au$ at RHIC}

To extend our study  to jets in heavy-ion collisions, we 
assume $\langle k_T^2\rangle\approx \mu^2$ (the Debye screening mass)
and a gluon density profile
$\rho(y)=(\tau_0/\tau)\theta(R_A-y)\rho_0$ for a 1-dimensional 
expanding system. Since the initial 
jet production rate is independent of the final gluon density 
which can be related to the 
parton-gluon scattering cross section \cite{bdms} [
$\alpha_s x_TG(x_T)\sim \mu^2\sigma_g$], one has then
\begin{equation}
\frac{\alpha_s T_{qg}^A(x_B,x_L)}{f_q^A(x_B)} \sim
\mu^2\int dy \sigma _g \rho(y)
[1-\cos(y/\tau_f)],
\end{equation}
where $\tau_f=2Ez(1-z)/\ell_T^2$ is the gluon formation time. One
can recover the form of energy loss in a thin plasma obtained 
in the opacity expansion approach \cite{glv},
\begin{equation}
\langle \frac{dE}{dL}\rangle \approx \frac{\pi C_aC_A\alpha_s^3}{R_A}
\int_{\tau_0}^{R_A} d\tau \rho(\tau) (\tau-\tau_0)\ln\frac{2E}{\tau\mu^2},
\end{equation}
where we have assumed $\sigma_g\approx C_a 2\pi\alpha_s^2/\mu^2$ 
($C_a$=1 for $qg$ and 9/4 for $gg$ scattering).

Neglecting the logarithmic dependence on $\tau$, the averaged energy loss
in a 1-dimensional expanding system can be expressed as \cite{ww02}
\begin{equation}
\langle\frac{dE}{dL}\rangle_{1d} \approx \frac{dE_0}{dL} \frac{2\tau_0}{R_A};
\;\;\; \frac{dE_0}{dL}=\frac{\pi C_aC_A\alpha_s^3}{2}\rho_0 R_A
\ln\frac{2E}{\tau_0\mu^2}
\end{equation}
where $dE_0/dL$
is the energy loss in a static medium with the same gluon density $\rho_0$ 
as in a 1-d expanding system at an initial time $\tau_0$.
Because of the expansion, the averaged energy loss $\langle dE/dL\rangle_{1d}$
is suppressed as compared to the static case and does not depend linearly
on the system size. Similarly, one have the total transverse momentum
broadening in an expanding system,
\begin{equation}
\langle \Delta q_\perp^2\rangle
=2\pi C_a\alpha_s^2\int_{\tau_0}^R d\tau\rho(\tau)
=\langle \Delta q_\perp^2\rangle_0
\frac{\tau_0}{R_A}\ln\frac{R_A}{\tau_0};
\;\;\; \langle \Delta q_\perp^2\rangle_0=2\pi C_a\alpha_s^2\rho_0
\end{equation}

Fitting the PHENIX data \cite{phenix}
on suppression of large $p_T$ $\pi_0$ spectra in Fig.~\ref{fig2}
yields  $\langle dE/dL\rangle_{1d} \approx 0.34\ln E/\ln 5$ GeV/fm.
Taking into account the expansion, this would be equivalent 
to $(dE/dL)_0=0.34 (R_A/2\tau_0)\ln E/\ln 5$ in a static system 
with the same gluon density as the initial value of the
expanding system at $\tau_0$. With $R_A\sim 6$ fm and
assuming $\tau_0\sim 0.2$ fm, 
this would give $(dE/dL)_0\approx 7.3$ GeV/fm for a 10-GeV parton, 
Since the parton energy loss is directly proportional to gluon density of
the medium, this implies that the gluon density in the initial stage 
of $Au+Au$
collisions at $\tau_0=0.2$ fm/$c$ is about 15 times higher than that 
inside a cold nucleus. To extract the parton density and the
initial formation time, one has to measure the energy loss and
$q_T$ broadening simultaneously.

In Fig.~\ref{fig2}, we also show the predicted suppression factor at
large $p_T$ with the same logarithmic energy-dependence of the 
energy loss as in
the cold nuclear matter which gives a suppression factor that increases
with $p_T$. However, new RHIC data \cite{phenix} at 200 GeV show
a constant suppression factor at larger $p_T$. This indicates
the importance of the detailed balance \cite{ww2} which gives
much stronger energy dependence of the energy loss. We show as the lower dashed line
in Fig.~\ref{fig2} the result
for an effective parton energy loss
\begin{equation}
\Delta E\propto (E/\mu-1.6)^{1.20}/(7.5+E/\mu)
\label{para}
\end{equation}
which is parameterized according to the result from Ref.~\cite{ww2}
where both stimulated gluon emission and thermal absorption are
included in the calculation of the total energy loss. The detailed
balance between emission and absorption reduces the effective parton
energy loss and on the other hand increases the energy dependence. 
The threshold is the consequence of gluon absorption that competes
with radiation that effectively shuts off the energy loss. The
parameter $\mu$ is set to be 1 GeV in the calculation shown in Fig.~\ref{fig2}.

This work was supported by DOE under Contract No. DE-AC03-76SF00098.

\end{document}